\documentclass[twocolumn]{aastex7}
\usepackage{comment}
\usepackage[flushleft]{threeparttable}
\usepackage{xcolor}
\usepackage{commath}
\usepackage{soul}
\usepackage{CJKutf8}
\usepackage[normalem]{ulem}

\def\simgr{\mathrel{\hbox{\rlap{\hbox{\lower4pt\hbox{$\sim$}}}\hbox{$>$}}}}
\newcommand{\mi}{M_\mathrm{i}}

\newcommand{\qi}{q_\mathrm{i}}

\newcommand{\porbi}{P_\mathrm{orb,i}}

\newcommand{\etaCE}{\eta_\mathrm{CE}}

\newcommand{\logpi}{\mathrm{log}\,P_\mathrm{orb,i}}

\newcommand{\dd}{\mathrm{d}}

\def\porbi{P_\mathrm{orb,i}}
\def\mso{M_\odot}

\def\rs{R_\odot}

\def\kms{\mathrm{km\,s}^{-1}}

\def\dd{{\mathrm d\,}}

\begin{document}
\begin{CJK*}{UTF8}{gkai}

\title{Tracing the Physical Lineage of GRB 211211A: Population Constraints on NS-WD Merger Gamma-Ray Bursts}

\correspondingauthor{Xiao-Tian Xu; Bin-Bin Zhang}
\email{xxu.astro@outlook.com;bbzhang@nju.edu.cn}

\author[orcid=0000-0001-9565-9462,sname='Xu']{Xiao-Tian Xu (徐啸天)}
\affiliation{School of Astronomy and Space Science, Nanjing University, Nanjing 210093, People{'}s Republic of China}
\affiliation{Argelander-Institut f\"ur Astronomie, Universit\"at Bonn, Auf dem H\"ugel 71, 53121 Bonn, Germany}
\email{xxu.astro@outlook.com; xxu@astro.uni-bonn.de}

\author[orcid=0000-0003-4111-5958,sname='Zhang']{Bin-Bin Zhang}
\affiliation{School of Astronomy and Space Science, Nanjing University, Nanjing 210093, People{'}s Republic of China}
\affiliation{Key Laboratory of Modern Astronomy and Astrophysics (Nanjing University), Ministry of Education, Nanjing 210023, People's Republic of China}
\email{bbzhang@nju.edu.cn}

\author[sname='Guo']{Yun-Lang Guo}
\affiliation{School of Astronomy and Space Science, Nanjing University, Nanjing 210093, People's Republic of China}
\affiliation{Key Laboratory of Modern Astronomy and Astrophysics (Nanjing University), Ministry of Education, Nanjing 210023, People{'}s Republic of China}
\email{yunlang@nju.edu.cn}

\author[sname='Li']{Xiang-Dong Li}
\affiliation{School of Astronomy and Space Science, Nanjing University, Nanjing 210093, People{'}s Republic of China}
\affiliation{Key Laboratory of Modern Astronomy and Astrophysics (Nanjing University), Ministry of Education, Nanjing 210023, People{'}s Republic of China}
\email{lixd@nju.edu.cn}
\begin{abstract}

The peculiar long gamma-ray burst (GRB) event, GRB 211211A, is known for it is association with a kilonova feature. Whereas most long GRBs are thought to originate in the core collapse of massive stars, the presence of kilonova suggests GRB 211211A was instead produced by a merger of a compact object binary. Building on the interpretation put forward by \citet{Yang2022Natur.612..232Y}--who argue that GRB 211211A was powered by a massive white-dwarf + neutron-star (WD-NS) merger--we adopt this WD-NS scenario as our observationally supported starting point. If the burst truly originates from that channel, its rarity must mirror the formation and merger rate of WD-NS binaries--a rate still largely unexplored in conventional massive-binary population studies. In this {work}, we present a qualitative analysis based on binary evolution physics in order to understand the fraction of GRB 211211A in short GRBs (NS-WD/NS-NS fraction).  Since the progenitors of massive WD-NS binaries occupy the initial mass function-preferred regime,
where the zero-age main-sequence mass range of the assumed WD mass range (1.2-1.4$\,M_\odot$) is comparable to that of NSs,
the NS-WD/NS-NS fraction emerging from our standard evolutionary path is expected to be $\sim$14--37\%,  far higher than the observed fraction ($\sim5$\%). This discrepancy might imply a large, still-unidentified population of GRB 211211A-like events or an unusual origin of the NS-such as being hypernova-born or accretion-induced-collapse-born. Placing these results in a broader compact-binary context, implications for black-hole systems are also discussed.
\end{abstract}


\received{}
\revised{}
\accepted{}
\submitjournal{ApJ}

\section{Introduction}

Short‐duration $\gamma$-ray bursts (SGRBs) deliver $\sim10^{52}\,$erg in $\gamma$-rays within $T_{90}\lesssim2$ s and are followed by luminous X-ray afterglows \citep{Zhang2006ApJ...642..354Z}. Their leading engine is the merger of compact binaries—BH, NS, or WD \citep{Woosley2006}—a link firmly established for the NS–NS channel by the joint GW170817/GRB 170817A detection \citep{Abbott2016}. Nonetheless, several recent bursts display properties that depart from this canonical picture \citep[e.g.,][]{Zhang2021NatAs...5..911Z,Yang2022Natur.612..232Y,Zhong2024ApJ...963L..26Z,Freeburn2025MNRAS.537.2061F,Geng2025ApJ...984L..65G}.

Recently a peculiar long GRB (duration above two seconds), GRB 211211A, was observed \citep{Yang2022Natur.612..232Y}. 
Its unique multi-wavelength emission makes GRB 211211A different from most of the long GRB events that are usually expected to  originate from dying massive stars \citep{Woosley2006}. Particularly, GRB 211211A was accompanied by a kilonova-like feature, which is powered by neutron-rich ejecta and typical in SGRBs \citep{Li1998ApJ...507L..59L,Abbott2017ApJ...848L..12A}. This observational fact implies the non-collapsar-origin nature of GRB 211211A.

To explain the nature of GRB 211211A, several scenarios have been proposed, including a black hole-neutron star (BH-NS) merger \citep{Rastinejad2022Natur.612..223R,Sarin10.1093/mnras/stac3441} and a white dwarf-neutron star (WD-NS) merger \citep{Yang2022Natur.612..232Y,Zhong2023ApJ...947L..21Z,Peng2024ApJ...967..156P}. In this {work}, we focus on the WD-NS merger model proposed by \citet{Yang2022Natur.612..232Y}, which posits that the strong gravitational field of the neutron star induces the neutralization of the white dwarf material, thus powering the kilonova. This mechanism requires the white dwarf to have a near-Chandrasekhar mass. Previous modelling by \citet{Zhong2023ApJ...947L..21Z} suggests that GRB 211211A-like events can arise as long as the white dwarf mass exceeds $1 M_{\odot}$. In addition to emission characteristics, the event rate of GRB 211211A provides further insight into its physical nature. While \citet{Yin2023ApJ...954L..17Y} has conducted an empirical estimate of the event rate for GRB 211211A-like events in the local Universe, this rate has rarely been discussed within the context of binary stellar evolution. Recently, \citet{Chrimes2025arXiv250810984C} has performed a population synthesis calculations, showing, in the local Universe, the fraction of GRB 211211A-like events is about 10\% of NS-NS mergers. In this work, we present an independent study and reach a similar result by using a different method.

Rather than pursuing a comprehensive population synthesis calculation, this {work} aims to offer a qualitative analysis grounded in the physics of massive binary evolution to understand the rarity of GRB 211211A. Following \citet{Yang2022Natur.612..232Y}, we examine the possibility that GRB 211211A originated from a WD-NS merger. First, in \S\ref{sect2}, we revisit the rarity of GRB 211211A, specifically its fraction within the broader population of SGRBs. In \S\ref{sect3}, we describe the evolutionary paths considered in this work and outline our qualitative approach. In \S\ref{sect4}, we present our results and discuss their implications for the NS-WD merger scenario. Finally, we summarize our conclusions in \S\ref{sect5}.

\section{Revisit the rarity of GRB~211211A}\label{sect2}

The population of confirmed Type~I GRBs compiled by \citet{Yang2022Natur.612..232Y} currently numbers \(N_{\mathrm{I}} = 41\).%
\footnote{Throughout this {work} we adopt the phenomenological ``Type~I / Type~II'' classification introduced by \citet{Zhang2009ApJ...703.1696Z}, where Type~I events are compact-object mergers and Type~II events are collapsars.}
Within this sample, three bursts possess \(\mathrm{T}_{90}>3\ \mathrm{s}\), markedly longer than the canonical short-burst threshold:

\begin{itemize}
 \item GRB~060614 (\(\mathrm{T}_{90}\simeq 102\ \mathrm{s}\); \citealt{Gehrels2006}),
 \item GRB~211211A (\(\mathrm{T}_{90}\simeq 43.18 \ \mathrm{s}\); \citealt{Yang2022Natur.612..232Y}),
 \item GRB~230307A (\(\mathrm{T}_{90}\simeq 41.52\ \mathrm{s}\); \citealt{Sun2024}).
\end{itemize}

While GRB~060614 can be interpreted as a short, hard spike followed by an extended soft tail, the other two long--$\mathrm{T}_{90}$ events---GRB~211211A and GRB~230307A---show \emph{definitive} hard $\gamma$-ray emission throughout their prompt phases. Their long durations therefore reflect long-term engine activity, not merely a faint soft tail. Both bursts also display kilonova signatures, extended emission, and other features that are difficult to accommodate within the standard neutron-star binary (NS--NS) merger channel \citep{Yang2022Natur.612..232Y}. These anomalies point to an alternative central engine, most plausibly the merger of a neutron star with a massive white dwarf (NS--WD; e.g.\ \citealt{Zhong2023ApJ...947L..21Z}). Further support comes from \citet{Sun2024}, who infer a magnetar remnant for GRB~230307A---a natural outcome of an NS--WD merger \citep{Yang2022Natur.612..232Y}. {The recent detection of gamma-ray periodicity in GRB~230307A is also in consistent with a central engine of a millisecond magnetar \citep{Chen2025NatAs.tmp..194C}.}

If we exclude GRBs~211211A and 230307A from the Type~I sample, the remaining population plausibly traces NS--NS (or, more rarely, NS--BH) mergers, giving 
\(N_{\mathrm{NS\text{--}NS}} = 41 - 2 = 39\). 
We may then define the empirical fraction of NS--WD to NS--NS merger GRBs as
\begin{equation}
\begin{split}
f_{\mathrm{NS\text{--}WD,obs}} \;&=\;
\frac{N_{\mathrm{NS\text{--}WD}}}{N_{\mathrm{NS\text{--}NS}}}
\;=\;
\frac{2_{-1.3}^{+2.6}}{39_{-6.3}^{+7.3}}\\
\;&=\;
0.0513_{-0.0299}^{+0.0480}
\;\;(\simeq 5.1_{-3.0}^{+4.8}~\%),
\end{split}
 \label{eq:fraction}
\end{equation}
{where the 1-$\sigma$ confidence interval is estimated by the Poisson error \citep{Gehrels1986ApJ...303..336G}. We note that it is unclear how many GRB~211211A-like events remain unidentified in the data, and this empirical fraction may need revision depending on the future detections of more GRB~211211A-like events.}

Equation~\eqref{eq:fraction} gives an observational reference point for the intrinsic merger-rate ratio \(\mathcal{R}_{\mathrm{NS\text{--}WD}}/\mathcal{R}_{\mathrm{NS\text{--}NS}}\). In the following sections we confront this 5\% benchmark with expectations from binary-population estimate
(\S\ref{sect3}) and from the Galactic census of compact binaries (\S\ref{MW}). Any theoretical scenario that hopes to explain GRB~211211A must reconcile its extreme phenomenology with the modest, but non-negligible, frequency implied by Eq.~\eqref{eq:fraction}.

\section{Evolutionary paths and method\label{sect3}}

\subsection{{Evolutionary paths}}

\begin{figure*}[!thbp]
 \centering
 \includegraphics[width=\linewidth]{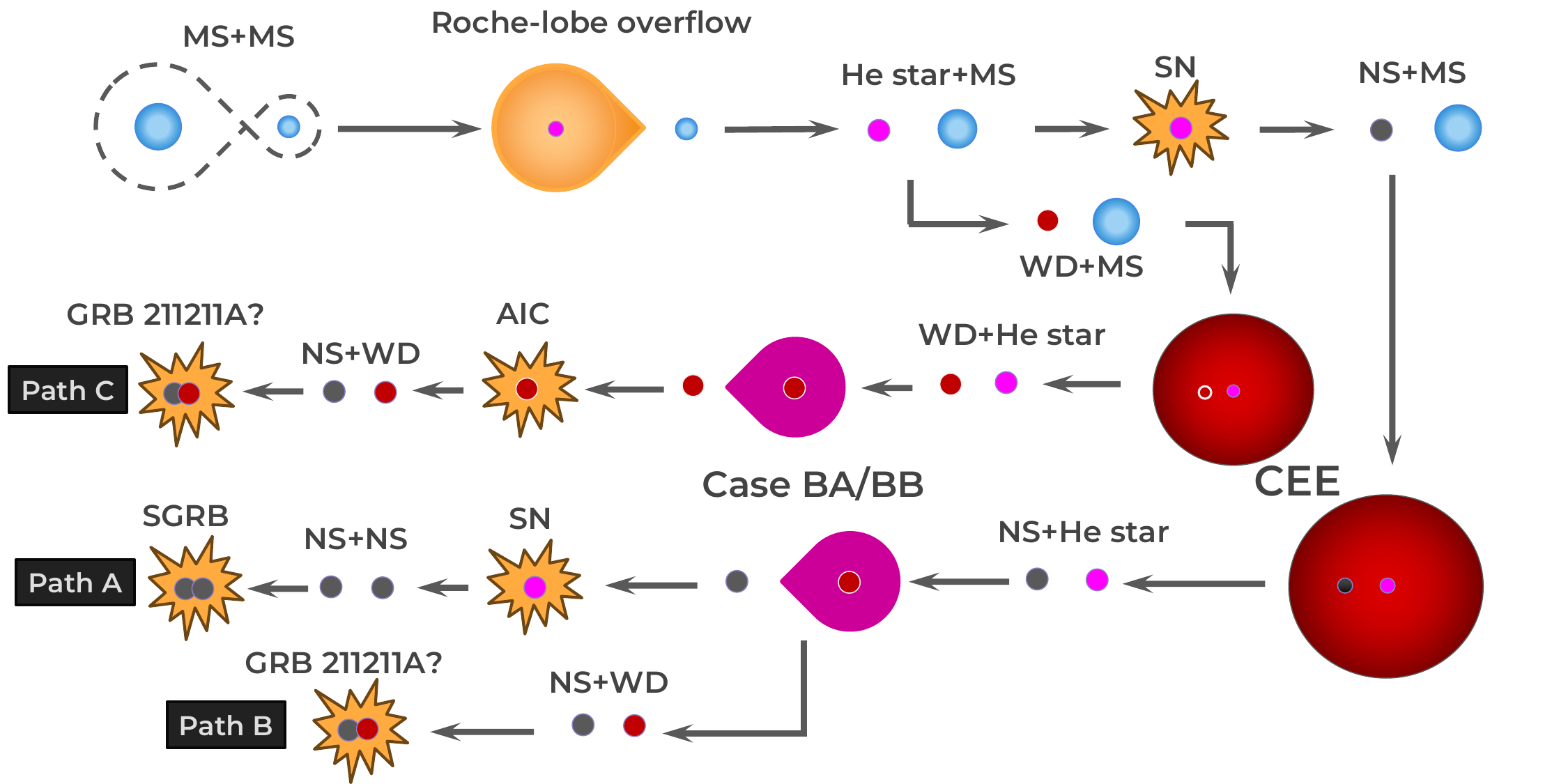}
 \caption{Schematic evolution of a massive binary towards to short $\gamma$-ray bursts (SGRBs; Path\,A) and GRB 211211A (Paths\,B and C). The meaning of the abbreviations are indicated in the following: 
 1) ``MS", main-sequence, 
 2) ``He star", helium star, 
 3) ``WD", white dwarf, 
 4) ``SN", supernova, 
 5) ``NS", neutron star, 
 6) ``CEE", common envelope evolution, 
 7) ``Case BA/BB", mass transfer with a core helium burning/depleted mass donor, 
 8) ``AIC", accretion-induced collapse, 
 and 9) ``SGRB", short gamma-ray burst. \label{schematic} 
 }
\end{figure*}

In this {work}, we consider three evolutionary paths of isolated massive binaries \citep[see][for comprehensive reviews]{Han2020RAA....20..161H,Marchant2023}, which are outlined in Fig.\,\ref{schematic}, where Path\,A for short $\gamma$-ray bursts originated from NS-NS mergers, and Paths\,B and C for GRB 211211A-like events. All paths start with a massive main-sequence (MS) binary, and then the expansion of the primary star triggers a mass transfer phase, which we expect to be stable (i.e., Roche-lobe overflow; RLO) in order to avoid merging during the later common envelope evolution (CEE; see below). After the RLO phase, the primary star leaves its naked helium core (He star), which may undergo supernova (SN) explosion (Paths\,A and B), producing a NS-MS binary if SN-induced disruption is avoided. Due to the extreme mass ratio of the NS-MS binary, the mass transfer triggered by the  initial\footnote{In the {work}, ``initial" refers to the zero-age main-sequence (ZAMS) point.} secondary star should be unstable, leading to a common envelope phase \citep{Ivanova2013}, where the mass gainer spirals into the envelope of the giant mass donor. This process mostly results in a merger. A CEE survivor is expected to be a NS-He star binary. After a potential mass transfer phase with a core helium burning/depleted mass donor \citep[Case BA/BB; ][]{Dewi2002MNRAS.331.1027D}, the system forms a NS-NS system (Path\,A) or a WD-NS system (Path\,B), the progenitors of SGRBs \citep{Abbott2017ApJ...848L..12A} or GRB\,211211A \citep{Yang2022Natur.612..232Y}. For Path\,C, the initial primary star evolves to a WD without experiencing a SN event. In further evolution, it may become a NS through accretion-induced collapse due to the mass accretion during the Case BA/BB mass transfer \citep{Kato2004ApJ...613L.129K,Piersanti2014MNRAS.445.3239P,Wang2017MNRAS.472.1593W}, and then the initial secondary star produces a massive WD, forming a GRB 211211A progenitor.

\subsection{{Population estimate method}}

\def\mwdl{M_\mathrm{WD,1}}
\def\mwdu{M_\mathrm{WD,2}}
\def\mnsi{M_\mathrm{NS,i}}
\def\mbhi{M_\mathrm{BH,i}}
\def\etaWD{\eta_\mathrm{WD}}
\def\etaNS{\eta_\mathrm{NS}}

{In this subsection, we introduce our qualitative method, which we refer to as the population estimate method. The event rates expected from the evolution of isolated binaries are determined by two factors, the initial parameter space of the progenitor systems (i.e, initial primary mass, mass ratio, and orbital period), and the details of binary evolution (\S\ref{binary_physics}). Our method focuses on the first factor, and we introduce one parameter to reflect the effects of binary evolution.}
In order to measure the rarity of GRB 211211A emerging from Path\,B (the standard case), we introduce a ratio factor $f_\mathrm{NS-WD}$, defined as 
\begin{equation}
\begin{split}
 &f_{\rm NS-WD} = \frac{\eta_{\rm WD}\int_{\mnsi}^{\mbhi}\mi^{-\alpha}\int_{\mwdl/M_{\rm i}}^{\rm\mwdu/M_{\rm i}}\,\dd \qi\,\dd \mi}{\eta_{\rm NS}\int_{\mnsi}^{\mbhi}\mi^{-\alpha}\int_{\mwdu/M_{\rm i}}^1\,\dd \qi\,\dd \mi}, 
\label{f_NSWD}
\end{split}
\end{equation}
where $\qi$ is the initial mass ratio (secondary/primary), $\mi$ is the initial primary mass, which we assume to follow the initial mass function (IMF) $\mi^{-\alpha}$ \citep[$\alpha=2.3$;][]{Kroupa2001}, $M_\mathrm{WD,1/2}$ is the lower/upper ZAMS mass of forming a massive WD, $\mnsi$ is the NS/WD mass boundary at ZAMS, $\mbhi$ is the BH/NS mass boundary at ZAMS, and $(\eta_{\rm WD}/\eta_{\rm NS})$ is a correction factor determined by the details of massive binary evolution. 
The numerator and denominator of Eq\,\eqref{f_NSWD} integrate the initial parameter space of the assumed progenitors of GRB 211211A and SGRBs respectively.
This equation assumes all the stars are born at the same time (i.e., star burst). Since the event delay time distribution of GRBs is highly uncertain, it is unclear whether the observed type I GRB population is closer to the outcome of a star burst or a constant star formation. In \S\ref{CSFR}, we show that the uncertainty in star formation history does not affect our estimate.

In addition, we simply assume flat distributions for initial mass ratio and orbital period distributions. The reason is that the mass ratio distribution is observed to be near flat \citep{Sana2012}, and we expect merging NS-NS/WD binaries to emerge from a similar initial orbital period range (See \S\ref{general_fNSWD} for a general definition of $f_{\rm NS-WD}$).

We estimate the value of $(\etaWD/\etaNS)$ by examining the key process along massive binary evolution. As SGRBs are observed at cosmological distance \citep[e.g.,][]{Sun2015ApJ...812...33S}, their progenitors were likely born in low-metallicity environment. We rely on the detailed stellar evolution models in \citet{Wang2020} and \citet{Xu2025arXiv250323876X} to estimate the evolution of massive stars, which are computed at the metallicity of the Small Magellanic Cloud, $\sim$1/5th of the solar value (see \S\ref{stellar_models} for fitting formulas).
While the correction factor $(\etaWD/\etaNS)$ should depends on the initial primary mass, we only use the binaries with $\mi=10\mso$ and an initial orbital period of 100\,d to estimate $(\etaWD/\etaNS)$ (see \S\ref{binary_physics} for the assumptions on binary evolution physics). The considered stellar population is dominated by the low-mass end (i.e., $\mi\sim10\mso$) due to the IMF, and 100\,d is close to the orbital period boundary of stable/unstable mass transfer \citep[see Fig.\,A.1 in][]{Xu2025arXiv250323876X}. We estimate $(\etaWD/\etaNS)$ to be about 0.5--1 for NS-NS from Path\,A and WD-NS from Path\,B (see \S\ref{estiamte_eta} for details). 

\subsection{{Comparison between population synthesis and population estimate}}

{While population synthesis is a powerful tool in understanding the mergers of compact objects \citep[e.g.,][]{Belczynski2002ApJ...571..394B,Belczynski2008}, the calculation is subject to many sensitive factors, like mass transfer efficiency, mass transfer stability, and the efficiency of ejecting envelope during a common envelope phase. Those uncertainties are captured by a single parameter, $(\eta_{\rm WD}/\eta_{\rm NS})$, in our population estimate method, which can be calibrated by population synthesis or observations. It also makes our calculation more reproducible than population synthesis, and allows us to better focus on the initial parameter space of the progenitors of WD-NS binaries, which is evaluated by the relatively well-understood relation between the stellar mass at zero-age main sequence and the carbon-oxygen core mass at core carbon depletion (see \S\ref{stellar_models} and \S\ref{binary_physics} for details).
Another advantage of this method is that it gives a similar event-rate ratio as population synthesis calculations without involving massive Monte Carlo simulations (\S\ref{sect4}), making it an efficient approach for understanding the observed event-rate ratio. We note that our method has strong limitations that we are unable to predict distribution functions, which has to be derived by using detailed population synthesis calculations.
}

\section{Results: the fraction of GRB 211211A-like events in short Gamma-ray bursts\label{sect4}}

\begin{figure}[!t]
 \centering
 \includegraphics[width=\linewidth]{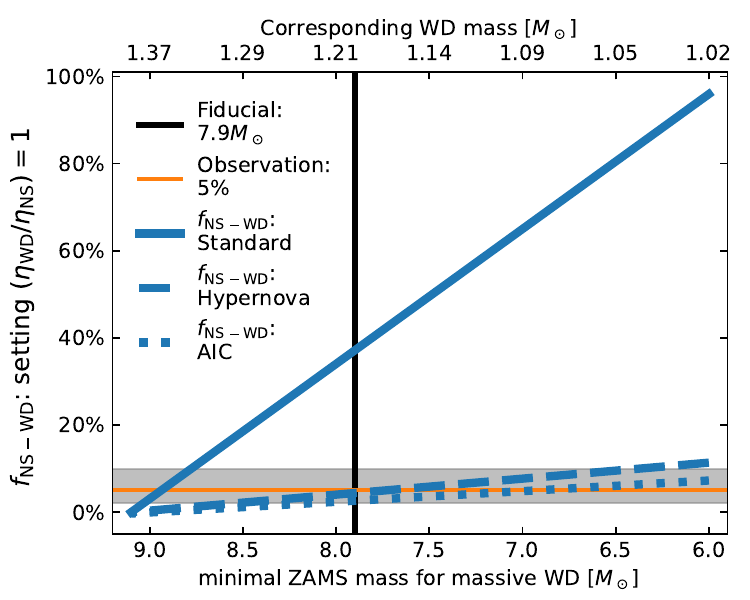}
 \caption{The calculated fractions of neutron star-white dwarf (NS-WD) binaries in NS-NS binaries as the function of the minimal zero-age main-sequence mass assumed for the WD. The corresponding WD masses are indicated on the top. The solid blue line, dashed blue line, and dotted blue line correspond to three evolutionary paths for the NS-WD binaries, which are the Standard case (Eq.\,\ref{f_NSWD_pathB}; Path\,B in Fig.\,\ref{schematic}), the hypernova case (Eq.\,\ref{f_NSWD_hyper}; Path\,B but at high-mass regime), and the accretion-induced collapse case (Eq.\,\ref{f_NSWD_AIC}; Path\,C in Fig.\,\ref{schematic}). In addition, the black vertical line indicates our fiducial value, and the horizontal orange line marks the 5\% benchmark from observations{, with the grey region indicating the confidence interval estimated by the Poisson error} (Eq.\,\ref{eq:fraction}). }
 \label{fig:all_f_NSWD}
\end{figure}

\subsection{Constraints from the mass of white dwarf\label{result1}}

We firstly focus on Paths\,A and B. 
We take the typical value, $1.4\mso$\footnote{ While, observationally, NS mass exhibits diverse values, our choice of $1.4\mso$ does not change the qualitative conclusion in this {work}.}, for the NS mass, leading to a ZAMS NS/WD boundary of 9.1$\mso$, below which WD formation is expected. For the BH/NS boundary, we take 20$\mso$, which is related to the sudden increase in compactness parameter of pre-collapse stars \citep{Sukhbold2018}.  For the WD mass range, 
we consider a narrow range for WD mass 1.2$-$1.4$\mso$ (7.9$-$9.1$\,\mso$ at ZAMS; see \S\ref{stellar_models} for formulas) as our fiducial range, which we assume to allow material neutralisation during merger as suggested in \citet{Yang2022Natur.612..232Y}.

Figure\,\ref{fig:all_f_NSWD} presents the computed $f_{\rm NS-WD}$ (Standard; Eq.\,\ref{f_NSWD}), as a function of $M_{\rm WD,1}$, i.e., the lower ZAMS mass of forming a massive WD, where the hypernova case and AIC case will be introduced in \S\ref{result_hypernova} and \S\ref{result_AIC} respectively, and $(\etaWD/\etaNS)$ is taken to be unit for presenting purpose. As expected, the $f_{\rm NS-WD}$ increases rapidly with decreasing $\mwdl$ (i.e., wider WD mass range). Our fiducial mass range gives
\begin{equation}
 f_\mathrm{NS-WD} = 37\% \left(\frac{\etaWD}{\etaNS}\right)_{\rm Standard}\sim 14\text{--}37\%.
 \label{f_NSWD_pathB}
\end{equation}
The estimated fraction is significantly affected by the population in the IMF-preferred low-mass end. 
A given WD mass window corresponds to a much wider ZAMS mass window, making the parameter space for forming massive WD-NS and NS-NS systems comparable in the IMF-preferred regime. 
For example, with $\mi=10\mso$, whether the systems produce a NS-NS or WD-NS depends on the mass of the secondary. The ZAMS mass window for the secondary star to form a NS  is $[9.1,\,10]\,\mso$ ($0.9\mso$ width), and $[7.9,\,9.1]\,\mso$ (1.2$\mso$ width) for forming a 1.2$-$1.4$\,\mso$ WD, resulting in a NS/WD ratio close to unit at $\mi=10\mso$.

Our estimate is also qualitatively consistent with previous population synthesis calculations and estimates \citep{Belczynski2002ApJ...571..394B,Kim2003ApJ...584..985K,Kim2004ApJ...616.1109K,Eldridge2011MNRAS.414.3501E,Shao2018ApJ...867..124S,Toonen2018A&A...619A..53T,Eldridge2019MNRAS.482..870E,Grunthal2021MNRAS.507.5658G,He2024MNRAS.529.1886H,Biswas2025arXiv250611676B,Chrimes2025arXiv250810984C}. 
For instance, adopting a threshold WD mass of 1.25$\,\mso$ that allows NS-WD mergers, \citet{He2024MNRAS.529.1886H} obtained a galactic merger rate of 5-30\,Myr$^{-1}$, while the merger rate of galactic NS-NS systems is estimated to be 32$_{-9}^{+19}\,$Myr$^{-1}$ based on the observed sample \citep{Grunthal2021MNRAS.507.5658G}. Furthermore, the recent BPASS population synthesis calculation by \citet{Chrimes2025arXiv250810984C} has also found $f_{\rm NS-WD}\sim10\%$ in the local Universe if a mass ratio (WD/NS) of above 0.9 is assumed for NS-WD systems.

However, our estimated fraction is much higher than the observed 5\% benchmark (Fig.\,\ref{fig:all_f_NSWD}).
This discrepancy might imply that there exists a large population of unidentified GRB 211211A-like events, the mergers of massive WD-NS binaries are abundant but mainly ends up with other types of transients, like type Ia SNe, or the mass transfer from a massive WD towards a NS is stable, which is gentle process and cannot launch explosive transit events. 
Reproducing the observed 5\% though our standard path requires a WD mass above about 1.36$\mso$ (Fig.\,\ref{fig:all_f_NSWD}), which largely overlaps with the WD mass range that allows the occurrence of electro-capture reaction \citep{Nomoto1984}.  
As it is unclear whether such extreme WD exists or not, we further propose that the rarity of GRB 211211A might imply that the NS involved in the merger has a strong magnetic field, which slows down the accretion onto the NS, resulting in a long GRB duration.

In the following subsections, we further explore the possibilities that the NS involved in GRB 211211A was formed from unusual evolutionary paths that might be relevant to the formation of magnetars. We will focus on the corresponding $f_{\rm NS-WD}$ value. A discussion on the launching mechanism of GRB 211211A is beyond the scope of this {work}.

\subsection{Neutron star formed from a hypernova?\label{result_hypernova}}

Hypernovae are expected to be produced by stars with ZAMS masses above $\sim25\mso$ \citep{Woosley2006ARA&A..44..507W,Janka2012}, which is above the compactness peak at 20$\mso$ \citep{Sukhbold2018}. The outcome compact object could be a BH \citep{MacFadyen1999ApJ...524..262M} or a fast rotating magnetar \citep{Kasen2010ApJ...717..245K}. 
For GRB 211211A, we assume that the pre-merger NS was a magnetar born from a hypernova. When the merger occurred, the magnetic field of the NS had significantly decayed but might still be stronger than most of the NSs.
Meanwhile, the massive WD provides neutron-rich ejecta to power the kilonova as suggested by \citep{Yang2022Natur.612..232Y}. In the framework, the progenitor binary of GRB 211211A has also evolved through Path\,B in Fig\,\ref{schematic} but at a much higher mass range than that assumed in \S\ref{result1}. Then the corresponding
$f_{\rm NS-WD}$ fraction is estimated as 
\begin{equation}
\begin{split}
 &f_{\rm NS-WD} = \frac{\eta_{\rm WD}\int_{M_{\rm NS,hyper}}^{M_{\rm BH,hyper}}\mi^{-\alpha}\int_{\mwdl/M_{\rm i}}^{\rm\mwdu/M_{\rm i}}\,\dd \qi\,\dd \mi}{\eta_{\rm NS}\int_{\mnsi}^{\mbhi}\mi^{-\alpha}\int_{\mwdu/M_{\rm i}}^1\,\dd \qi\,\dd \mi}, \label{f_NSWD_hyper}
\end{split}
\end{equation}
where $M_{\rm NS,hyper}$ and $M_{\rm BH,hyper}$ are the lower and upper limit of the ZAMS mass window for hypernova-born NSs, which are taken to be 20$\mso$ and 30$\mso$ respectively (See \S\ref{CSFR} for the constant star formation case). The other terms remain the same values as Eq.\,\eqref{f_NSWD_pathB}. With our fiducial WD mass range, 1.2-1.4$\mso$, this hypernova scenario gives
\begin{equation}
 (f_{\rm NS-WD})_{\rm hyper} = 4.4 \% \left(\frac{\etaWD}{\etaNS}\right)_{\rm hyper}\sim 2.2\text{--}4.4\%,
\end{equation}
which is consistent with the observed fraction of GRB\,211211A. For a wider WD mass range, $(f_{\rm NS-WD})_{\rm hyper}$ increases up to about 10\% (Fig.\,\ref{fig:all_f_NSWD}), still consistent with the observed fraction. The estimated fraction could be largely reduced if hypernova has a stronger natal kick than normal supernova.

\subsection{Neutron star formed from accretion-induced collapse?\label{result_AIC}}

A WD can gain mass over the Chandrasekhar limit via accretion, then collapsing into a NS, which is called accretion-induced collapse \citep[AIC;][]{Nomoto1991ApJ...367L..19N,Dessart2006ApJ...644.1063D,Wang2020RAA....20..135W}. The outcome NS might rotate rapidly due to angular momentum accretion \citep{Dessart2006ApJ...644.1063D}. Usually, massive WDs are magnetised. Assuming the conversation of magnetic flux, an AIC-born NS might have magnetic fields stronger than normal NSs \citep{meng2025eccentricmillisecondpulsar}. For GRB\,211211A, if the NS was born from an AIC event, this process should occur during a Case BA/BB mass transfer (see Path\,C in Fig.\,\ref{schematic}). The corresponding $f_{\rm NS-WD}$ fraction is estimated to be
\begin{equation}
\begin{split}
 &f_{\rm NS-WD} = \left(\frac{\etaWD}{\etaNS}\right) \times\\
 &\frac{\int_{M_{\rm WD,AIC}}^{\mnsi}\mi^{-\alpha}\int_{M_{\rm WD,AIC}/M_{\rm i}}^{1}\,\dd \qi\,\dd \mi}{\int_{\mnsi}^{\mbhi}\mi^{-\alpha}\int_{\mwdu/M_{\rm i}}^1\,\dd \qi\,\dd \mi}, \label{f_NSWD_AIC} 
\end{split}
\end{equation}
where $M_{\rm WD,AIC}$ is the minimal ZAMS primary mass that allows AIC to occur (See \S\ref{CSFR} for the constant star formation case).

To obtain $M_{\rm WD,AIC}$, we need to estimate how much helium that a WD can accrete during a Case BA/BB mass transfer, which is highly uncertain. \citet{Wang2017MNRAS.472.1593W} show that a helium-accreting WD is expected to be ignited off-centre if the mass accretion rate is about $2\text{--}6\times10^{-6}\,\mso\,{\rm yr}^{-1}$, which leads to an AIC event. Otherwise, the WD would be ignited to explode as a Type Ia SN, or the accreted material accumulates on the surface of the WD, forming a giant-like configuration. For a qualitative estimate, we simply assume that a massive WD can accrete helium at $10^{-6}\,\mso\,{\rm yr}^{-1}$. We expect that most of the mass is transferred within the thermal timescale $\tau_{\rm th}$ of the mass donor, which is\footnote{See Sect. 2.4.1 ``The thermal timescale" in the lecture note by Pols: \url{https://www.astro.ru.nl/~onnop/education/stev_utrecht_notes/}.} 
\begin{equation}
 \tau_{\rm th} \approx 1.5\times 10^{7} \left(\frac{M}{\mso}\right)^2 \frac{R_\odot}{R} \frac{L_\odot}{L}\,{\rm yr},
\end{equation}
where $M$ is stellar mass, $R$ is stellar radius, and $L$ is stellar luminosity.
For a $3\,\mso$ helium star at the middle of core helium burning at the SMC metallicity, it has $R=0.26\,R_\odot$ and $L=10^4 \,L_\odot$, resulting in $\tau_{\rm th}\sim 0.5\times10^{5}{\rm\, yr}$. Then the amount of helium that a massive WD might accrete $\Delta M$ is 
\begin{equation}
 \Delta M \sim \tau_{\rm th} \times 10^{-6}\,\mso\,{\rm yr}^{-1} \sim 0.05\,M_\odot.
\end{equation}
Therefore, this scenario may require the initial primary star to form a WD with mass above 1.35$\,\mso$ (8.85$\,\mso$ at ZAMS). This estimate is consistent with the parameter space obtained by \citet{Liu2018MNRAS.477..384L}, who also found that, with a $3\mso$ helium star mass donor, the occurrence of AIC in close binaries requires the WD mass gainer to have a mass of 1.3$\mso$. Then, keeping the fiducial WD mass range for the initial secondary star, Eq.\,\eqref{f_NSWD_AIC} gives
\begin{equation}
 (f_{\rm NS-WD})_{\rm AIC} = 2.62\% \left(\frac{\etaWD}{\etaNS}\right)_{\rm AIC}.
 \label{f_NSWD_AIC_result}
\end{equation}
Different from Paths\,A and B, this AIC-involved evolution (Path\,C in Fig.\,\ref{schematic}) does not experience any supernova events, leading to a much higher value for $(\etaWD/\etaNS)$, which seems to overpredict the fraction of GRB\,211211A-like events. The mass accumulation efficiency related to helium burning is assumed to be unit. With a lower mass accumulation efficiency \citep[e.g.,][]{Kato2004ApJ...613L.129K}, the WD that undergoes AIC would be required to have a narrower mass range. Furthermore, if the Eddington accretion rate is below the threshold rate for AIC \citep{Piersanti2014MNRAS.445.3239P,Wang2017MNRAS.472.1593W}, helium-accreting massive WDs would end up with type Ia SNe. While the AIC scenario is highly uncertain, it may still reproduce the observed fraction of GRB 211211A.

\section{{Discussion: central engine}}

{In this work, we have focused on a specific scenario for GRB~211211A-like events, which is the merger of a NS and a massive WD, suggested by \citet{Yang2022Natur.612..232Y}. The merger product is expected to be a millisecond magnetar, whose magnetic dipole spin-down energy powers a Poynting-flux-dominated jet. The key requirement of this scenario is that the WD should be massive enough for triggering material neutralisation during merger to produce the observed kilonova feature, and it may also require the pre-merger NS to have a strong magnetic field as we have discussed in \S\ref{result_hypernova} and \S\ref{result_AIC}. In addition, past hydrodynamic simulations show that the mass transfer from a WD to a NS is unstable \citep{Bobrick2017MNRAS.467.3556B}, and the merger event may look similar to a SN Iax \citep{Bobrick2022MNRAS.510.3758B}. Those findings do not contradict with what we have proposed, since those simulations do not take into account magnetic fields. Those NS-WD mergers with a normally magnetised NS may produce a transit similar to SNe Iax \citep{Bobrick2022MNRAS.510.3758B}, but those with a strongly magnetised NS may produce a GRB~211211A-like event (\S\ref{result_hypernova} and \S\ref{result_AIC}). 
}

{However, from the perspective of population estimate or population synthesis, other scenarios, like the merger of a BH and a WD \citep{Lloyd-Ronning2024MNRAS.535.2800L} or a NS \citep{Rastinejad2022Natur.612..223R,Sarin10.1093/mnras/stac3441}, cannot be excluded theoretically. Those mergers may also achieve a event rate ratio of about 5\%, depending on the assumed binary physics \citep{Fryer1999ApJ...526..152F,Belczynski2002ApJ...571..394B,Mapelli2018MNRAS.479.4391M}. In the case of BH-WD/NS mergers, the BH is expected to launch a jet through the Blandford-Znajek mechanism, which depends on the BH spin and the accreted magnetic flux  \citep{Tchekhovskoy2010ApJ...711...50T,Wu2025ApJ...980L..28W}. Since the vast majority of stellar-born BHs are expected to have small natal spins due to the Tayler-Spruit dynamo \citep{Qin2019,Fuller2019ApJ...881L...1F}, it would require a highly magnetised companion to the BH, which does not contradict with a massive WD companion or a NS companion.
}

{One approach to distinguish a millisecond magnetar engine and a BH engine is to identify (quasi-)periodic signals in GRB events, as a millisecond magnetar is fast-rotating, and can potentially modulate the emission of the explosion. Recently, \citet{Chen2025NatAs.tmp..194C} report the detection of a potential $\gamma$-ray periodicity in GRB~230307A in a narrow time window, which seems to be consistent with a millisecond magnetar engine. Since GRB~230307A is a GRB~211211A-like event, it implies that GRB~211211A may also be powered by a millisecond magnetar, favouring the NS-WD merger scenario.}

\section{Conclusion\label{sect5}}

The  rarity of the peculiar event, GRB 211211A, should imprint its physical origin. In this {work}, we have revisited its rarity in terms of its fraction in short $\gamma$-ray bursts based on the type I GRB sample complied by \citet{Yang2022Natur.612..232Y}. 
Following \citet{Yang2022Natur.612..232Y}, we assume that GRB\,211211A was a merger of a NS and a massive WD, and we have performed a qualitative analysis in order to understand this rarity based on binary evolution physics. 

We found that it requires an extreme WD mass range to reproduce the observed fraction, 5\% (\S\ref{sect2}), though our standard evolutionary paths. 
This is because that, in the IMF-preferred low-mass end, the parameter space of forming NS-NS and assumed WD-NS systems are comparable, resulting in a fraction much higher than 5\%. This discrepancy might imply that there is a large population of unidentified GRB 211211A-like events, or most of the massive WD-NS mergers end up with other types of transits, like type Ia SN, instead of long GRBs.

We further propose that, besides the requirement for a massive WD, the NS might also be formed through unusual evolution paths, making GRB 211211A a rare event. We have explored two possibilities, hypernova and accretion-induced collapse, that might be relevant to the formation of magnetars. The corresponding fraction is roughly consistent with the observed fraction of GRB\,211211A. 

In addition, a similar analysis may also be applied to NS-NS, NS-BH, and BH-BH systems. Compared to BH-BH systems, the initial parameter space of forming NS-BH systems occupies the IMF-preferred regime. Considering the effects of supernova kicks, one might expect comparable intrinsic event rates for BH-NS/BH mergers or a higher event rate for NS-BH mergers. Compared to NS-NS systems, the progenitor binaries of BH-NS systems are more massive, i.e., more rare, but BH kicks, if exists, should be weaker than NS kicks \citep{Janka2017}, and BHs have more chance than NSs to survive CEE due to the difference in the masses of BHs and NSs. Which events are more abundant, NS-BH or BH-BH mergers, is highly model-dependent \citep[e.g.,][]{Mapelli2018MNRAS.479.4391M}.

\begin{acknowledgments}
{We thank the reviewer for the constructive feedback on the manuscript.}
We thanks Bing Zhang for insightful discussions.
We acknowledge the support by the National Key Research and Development Programs of China (2022YFF0711404, 2022SKA0130102, 2021YFA0718500), the National SKA Program of China (2022SKA0130100), the National Natural Science Foundation of China (grant Nos. 11833003, U2038105, U1831135, 12121003, 12041301, 12393811, 13001106, 12403035, 12573046), the science research grants from the China Manned Space Project with NO. CMS-CSST-2021-B11, the Fundamental Research Funds for the Central Universities.
\end{acknowledgments}

\bibliography{Xu_SMC}
\bibliographystyle{aasjournal}

\appendix

\section{Stellar models and fitting formulas\label{stellar_models}}

\def\lgmi{{\rm log\,}(M_{\rm i}/\mso)}
\begin{deluxetable*}{cc|cccc}
\tablecaption{Fitting to the stellar evolution models in \citet{Wang2020} and \citet{Xu2025arXiv250323876X}. In this table, variables $X$, $Y$, $a$, $b$, $c$, and $d$ have the same meaning as Eq.\,\eqref{fitting_formula}. \label{fitting}}
\tablehead{$Y$ & $X$ & a & b & c & d }
\startdata
$ M_{\rm He}/\mso$ & $\lgmi$ & -1.1286722 & 8.1477995 & -13.6530828 & 9.80763531\\
 $ R_{\rm He}/\rs$ & $\lgmi$ & 0.10802516 &-0.24784308 &0.43503492 & -0.06260567\\
 $\log R_{\rm MS}/\rs$ & $\lgmi$ & 0.02324015 & 0.56778424 & 0.06056826 & -0.03011988\\ 
 $ M_{\rm co}/\mso$ & $\lgmi$&-9.13133678 &39.51865733 &-52.33854403 & 23.54116\\
 $\log \lambda_{\rm bind}$&$\lgmi$& -2.16452866& 5.68288431& -8.3861172 &3.38944485\\
 \enddata
 \tablecomments{$M_{\rm He}$ is the helium core mass at core helium depletion, $R_{\rm He}$ is the radius of the helium at the middle of core helium burning, 
 $R_{\rm MS}$ is the stellar radius at core hydrogen mass fraction equal to 0.5, 
 $M_{\rm CO}$ is the CO core mass at core carbon depletion,
 $\lambda_{\rm bind}$ is the binding energy parameter at core helium depletion (see Sect.\,\ref{CEE}).
 }
\end{deluxetable*}

\def\lgmi{{\rm log\,}M_{\rm i}}

In order to estimate the evolution of massive binaries, we reply on the detailed stellar evolution models in \citet{Wang2020} and  \citet{Xu2025arXiv250323876X}, which were computed by the MESA code \citep{Paxton2011,Paxton2013,Paxton2015}, with the metallicity of the Small Magellanic Cloud. We adopt the following formula to fit the model data,
\begin{equation}
 Y = a + b \,X + c\,X^2 + d\,X^3,
 \label{fitting_formula}
\end{equation}
where $Y$ represents a model property, $X$ is the chosen independent variable, and $(a,\,b,\,c,\,d)$ are fitting parameters. Our fitting results are listed in Tab.\,\ref{fitting}.

\section{Binary evolution physics\label{binary_physics}}

In this section, we describe our assumption on binary evolution physics.

\subsection{Tides}

When the Roche-lobe filling factor of a star approaches unit, tidal torque increases significantly \citep[see][for equations of tidal torque]{Hurley2002}. We simply follow the assumption of \citet{Belczynski2002ApJ...571..394B} that all binaries are circularised when the mass donor fills the Roche lobe. The circularised orbital parameters are calculated by assuming the conservation of orbital angular momentum during the circularisation process.

\subsection{Roche-lobe overflow}

When a mass donor fills its Roche lobe, mass transfer occurs. We assume that the hydrogen-rich envelope of the donor star is completely stripped during the mass transfer, leaving a naked helium core, whose mass and radius are determined by the fitting formulas in \S\ref{stellar_models}. 

During the mass transfer, we assume a zero accretion efficiency due to the limitation of the critical rotation of the mass gainer \citep{Packet1981}. The non-accreted material is assumed to be ejected as isotropic wind, carrying away the specific orbital angular momentum of the mass gainer, during which the orbital evolution is calculated by Eq.\,(16.20) in \citet{Tauris2006}.

\subsection{Common envelope evolution\label{CEE}}

If mass transfer is unstable, the system enters common envelope (CE) evolution. We adopt the formalism of \citet{Webbink1984ApJ...277..355W} and \citet{deKool1990ApJ...358..189D} that the energy that drives the ejection of the mass donor's envelope is from the changing in orbital energy with an efficiency parameter $\etaCE$, which is taken to be unit. The binding energy of the envelope is measured by a parameter $\lambda_{\rm bind}$, which is given by the formula in \S\ref{stellar_models}, where only gravitational binding energy is considered. If the expected post-CE Roche lobe radius is smaller than the radius of the helium core of the donor star, we expect a merger during the common envelope evolution, where the Roche-lobe radius is determined by the fitting formula by \citet{Eggleton1983}. 

\subsection{Supernova explosion}

The mass ejection of supernova explosion is estimated by the mass difference between the pre-explosion helium star and the outcome neutron star, where the mass of the neutron star is fixed to be 1.4\,$\mso$ in this {work}. The corresponding kick velocities are described by a Maxwellian distribution with an 1D rms of $120\,\kms$ \citep{Kruckow2018}, which is adopted to describe a hydrogen-envelope-stripped explosion. We assume a random orientation for the direction of kicks. Whether a binary is disrupted or not depends on the orbital energy of the post-explosion system. If the orbital energy remains negative, the system survives the supernova event \citep[see detailed equations in][]{Hurley2002}. We calculate survival rates by performing Monte Carlo simulations for all pre-explosion systems. 

Supernova survivors could have significant eccentricities. If the separation at periastron is smaller than the radius of the main-sequence companion, we expect the binary to merge. 
The stellar radius is determined by the fitting formula in \S\ref{stellar_models}.

If the outcome compact object is a white dwarf, we assume zero momentum kick, and we use the CO core mass at core carbon depletion to determine the mass of the white dwarf.

\section{General definition of $f_{\rm NS-WD}$\label{general_fNSWD}}

Assuming initial primary mass $\mi$, initial mass ratio $\qi$, and logarithmic initial orbital period $\log\porbi$ following the initial mass function $\Phi_{\rm M}$, initial mass ratio function $\Phi_{\rm q}$, and initial orbital period function $\Phi_{\rm Porb}$, we define the fraction of NS-WD binaries in NS-NS binaries as
\begin{equation}
    f_{\rm NS-WD}=\frac{\int_{V_{\rm NS-WD}}\Phi_{\rm M}\Phi_{\rm q}\Phi_{\rm Porb}\,\dd\mi\,\dd\qi\,\dd(\log\porbi)}{\int_{V_{\rm NS-NS}}\Phi_{\rm M}\Phi_{\rm q}\Phi_{\rm Porb}\,\dd\mi\,\dd\qi\,\dd(\log\porbi)},
    \label{eq:f-NSWD-general}
\end{equation}
where $V_{\rm NS-WD}$ and $V_{\rm NS-NS}$ correspond to the parameter space of forming NS-WD and NS-NS systems. 
{In the main text, we have assumed a flat distribution for $q_{\rm i}$ and $\logpi$, which means $\Phi_{\rm q} = \Phi_{\rm Porb}= 1$. For simplicity, we have also assumed that the orbital period windows of the progenitors of NS-WD and NS-NS binaries share a similar width, $\Delta \logpi$. With these assumptions, Eq.\,\eqref{eq:f-NSWD-general} can be simplified as 
\begin{equation}
\begin{split}
    f_{\rm NS-WD}&=\frac{\int_{V_{\rm NS-WD}}\Phi_{\rm M}\,\dd\mi\,\dd\qi\,\dd(\log\porbi)}{\int_{V_{\rm NS-NS}}\Phi_{\rm M}\,\dd\mi\,\dd\qi\,\dd(\log\porbi)},\\
    &=\frac{ (\Delta \logpi)\iint_{S_{\rm NS-WD}}\Phi_{\rm M}\,\dd\mi\,\dd\qi}{(\Delta \logpi)\iint_{S_{\rm NS-NS}}\Phi_{\rm M}\,\dd\mi\,\dd\qi},\\
    &=\frac{\iint_{S_{\rm NS-WD}}\mi^{-\alpha}\,\dd\mi\,\dd\qi}{\iint_{S_{\rm NS-NS}}M^{-\alpha}\,\dd\mi\,\dd\qi},\\
\end{split}
\end{equation}
where $\Phi_{\rm M}$ is replaced by the initial mass function, $\mi^{-\alpha}$, $S_{\rm NS-NS}$ and $S_{\rm NS-WD}$ represent the complete parameter space of $\mi$ and $\qi$, which are shaped by the details of massive binary evolution. In the main text, we have adopted a parameter $(\etaWD/\etaNS)$ to reflect the effects of binary evolution on the initial parameter. 
}

\section{Estimation on the $(\etaWD/\etaNS)$ factor\label{estiamte_eta}}

\begin{deluxetable}{c|cccc|c}
\tablecaption{Survival rates of various evolutionary stages as an function of initial mass ratio $\qi$ (See \S\ref{estiamte_eta} for details). \label{eta_estimate}}
\tablehead{
 $\qi$ & \multicolumn{4}{c|}{Survival rate $\delta$ } &total\\
 & RLO &1st SN & CEE & 2nd SN & $\delta/\delta(\qi=0.9)$
 }
\startdata
 0.9 & 100\% & 4.22\% & 39.88\% & 100\% & 1\\
 0.8 & 100\% & 4.63\% & 42.29\% & 100\% & 1.16\\
 0.7 & 100\% & 5.38\% & 44.78\% & 100\% & 1.43\\ \hline
 \enddata
 \tablecomments{The estimate is based on the evolution of binaries having a 10$\mso$ initial primary star and a 100\,d initial orbital period with initial mass ratio from 0.9 to 0.7. We assume zero mass transfer efficiency, a Maxwellian velocity distribution with an 1D rms of 120\,$\kms$ for  the kicks of the first supernova event, while zero for the second supernova event, and one for the efficiency of ejecting envelope during a common envelope evolution. Stellar parameters are given by the fitting formula in \S\ref{stellar_models}, which is based on the detailed models in \citet{Xu2025arXiv250323876X}.}
\end{deluxetable}

Due to stellar mergers and supernova-induced disruption, the number of massive binary stars is expected to decrease significantly with binary evolution. The survival rate, which is the fraction of remaining systems after a given evolutionary stage, depends on various factors.  In the following, we estimate the survival rates associated with different evolution stages in order to estimate the value of $(\etaWD/\etaNS)$. We consider an initial mass ratio range of 0.7--0.9, covering our fiducial mass window for WD formation. The results are summarised in Tab.\,\ref{eta_estimate}. 

During the RLO phase, we assume that MS mass gainer cannot obtain mass (i.e., zero mass transfer efficiency), since the accretion onto a MS accretor in wide binaries is strongly limited by its critical rotation \citep{Packet1981,Wang2020,Langer2020,Rocha2024ApJ...971..133R,Xu2025arXiv250323876X}. All systems survive the first mass transfer phase under this assumption, while a higher mass transfer efficiency may lead to a large fraction of mergers \citep{Henneco2024,Schurmann2025arXiv250323878S}.

With our Monte Carlo simulations (see \S\ref{binary_physics} for our SN assumptions), we find that only about 4--5\% of the systems survive the SN explosion, which slightly decreases with initial mass ratio. This is because that, for a higher initial mass ratio, the mass ratio reversal occurs earlier during the mass transfer phase, resulting in a wider pre-SN orbit, which is easily disrupted.

We assume a merger during a CEE if the expected the post-CE Roche-lobe radius of the mass donor is smaller than the radius of its helium core. Hence this helium core radius corresponds to a minimal pre-CE orbital period for avoiding merging $P_{\rm crit,CE}$ (see \S\ref{binary_physics} for the CEE assumptions). 
With the post-SN systems generated from previous Monte Carlo simulations, we obtain a fraction of CEE survivors of about 40--45\% by counting the number of binaries with orbital periods below and above $P_{\rm crit,CE}$. The survival rate drops with increasing initial mass ratio is also due to the orbital widening during the RLO phase. For a tighter orbit, it has more chance to survive the SN event and reach the regime for CEE survivors.

\citet{Tauris2017} have shown that the observed galactic NS-NS binaries are consistent with ultra-stripped supernovae \citep[also see][]{Deng2024ApJ...963...80D}, which have small kick velocities and SN ejecta \citep{Tauris2015}. We accordingly assume zero supernova kicks for the second SN event. 
Consequently all the systems survive the second SN explosion. In addition, if we assume a considerable kick velocity, a large fraction of NS-NS systems would be disrupted. Therefore, our estimated $f_{\rm NS-WD}$ is a lower limit.

While the total survival rate decreases with initial mass ratio (the last column in Tab.\,\ref{eta_estimate}), it varies within a factor of 1--1.43 for initial mass ratio varying from 0.9 to 0.7. This means that the correction factor $(\etaWD/\etaNS)$ is close to unit. Since massive binary evolution is continuous with respect to the initial masses of stellar components, we expect this tendency and small variation to be also valid for higher initial primary masses. 

One major uncertainty of the above estimate is related to the merger delay time of NS-NS/WD binaries. If most of the massive WD-NS binaries have delay times close or above the Hubble time, while the NS-NS systems formed at the IMF-disfavoured regime have much shorter delay times, the correction factor $(\etaWD/\etaNS)$ would be much lower than unit. 
In order to address this uncertainty, in \S\ref{MW}, we have estimated the $(f_{\rm NS-WD})_{\rm MW}$ fraction of the observed galactic NS-NS/WD systems that have merger timescale below the Hubble time. We have found that a $(f_{\rm NS-WD})_{\rm MW}$ of 23\%. Hence we expect $(\etaWD/\etaNS)$ to be 0.5--1.

\section{Implications of galactic neutron star-neutron star/white dwarf sample\label{MW}}

In this section, we estimate the fraction of NS-WD mergers in NS-NS mergers by using the observed systems in the Milky Way, which are picked from the ATNF pulsar online catalogue\footnote{\url{https://www.atnf.csiro.au/research/pulsar/psrcat/}} \citep{Manchester2005AJ....129.1993M}. For WDs, we adopt the CO WDs with masses above $1\,M_\odot$, where CO WDs and ONeMg WDs are all classified as CO WDs in the ATNF catalogue. We fix the mass of the pulsar to be 1.4\,$\mso$, and the companion mass is taken to be the observed median mass (assuming a inclination of 60 degree). The merger delay time is computed by the equation in \citet{Mandel2021RNAAS...5..223M}, which captures the effect of eccentricity. We ignore the evolutionary time of the progenitor binaries, which is much lower than the merger timescale driven by gravitational wave radiation.

In order to estimate the event rate of NS-NS/WD mergers in the local Universe, we need to obtain the corresponding merger timescales as a function of redshift. \citet{Santoliquido2021MNRAS.502.4877S} and \citet{vanSon2025ApJ...979..209V} suggest that the formation of NS-NS systems is rather insensitive to metallicity. Accordingly, we simply assume that the NS-NS/WD systems at all redshifts have the same merger timescale distribution as the galactic population, which is enough for a crude estimate. The binaries that can merger at zero redshift should be born with a look-back time equal to its merger delay time, and the merger rate at zero redshift is proportional to the cosmic star formation rate density at the born-redshift of the binary (hereafter SFR weight). The relation between cosmic time and redshift is computed by the standard $\Lambda$CDM model with $(\Omega_{\rm M},\,\Omega_{\rm K},\,\Omega_{\rm \Lambda})=(0.3,\,0.0,\,0.7)$. We adopt the cosmic star formation history in \citet{Madau2014}.

Table\,\ref{SFR_weight} lists our computed merger delay time $\tau_{\rm M}$ and the corresponding SFR weights, we found the ratio of the total SFR weights of NS-WD to NS-NS is about 23\%, suggesting the event rates of NS-WD merger is about 23\% of NS-NS mergers at zero redshift. This estimate is about half of the fraction emerging from our initial parameter space estimate, Eq.\,\eqref{f_NSWD}, leading to a correction factor $(\etaWD/\etaNS)$ in 0.5--1. 

\begin{deluxetable}{ccccccc}
\tablecaption{Galactic neutron star-neutron star (NS-NS) and neutron star-white dwarf (NS-WD) binaries, and their SFR weights\label{SFR_weight}}
\tablehead{Name& $M_{\rm comp}$& $P_{\rm orb}$ & $e$ & $\tau_{\rm M}$ & $z_{\rm birth}$ & SFR weight\\
 & $(\mso)$ & (day) & & (Gyrs) & & $(\mso$\,Mpc$^{-3}$\,yr$^{-1})$}
\startdata
\multicolumn{6}{l}{NS-NS systems}\\
J0509+3801 & 0.761 & 0.380 & 0.59 & 0.99 & 0.059 & 1.74$\times10^{-2}$ \\
J0737-3039A & 1.559 & 0.102 & 0.09 & 0.07 & 0.004 & 1.51$\times10^{-2}$ \\
J0737-3039B & 1.739 & 0.102 & 0.09 & 0.06 & 0.004 & 1.51$\times10^{-2}$ \\
B1534+12 & 1.624 & 0.421 & 0.27 & 2.27 & 0.148 & 2.16$\times10^{-2}$ \\
J1756-2251 & 1.353 & 0.320 & 0.18 & 1.48 & 0.091 & 1.89$\times10^{-2}$ \\
J1757-1854 & 1.737 & 0.183 & 0.61 & 0.06 & 0.004 & 1.51$\times10^{-2}$ \\
J1906+0746 & 0.976 & 0.166 & 0.09 & 0.37 & 0.021 & 1.58$\times10^{-2}$ \\
J1913+1102 & 1.072 & 0.206 & 0.09 & 0.61 & 0.036 & 1.64$\times10^{-2}$ \\
B1913+16 & 1.056 & 0.323 & 0.62 & 0.39 & 0.023 & 1.59$\times10^{-2}$ \\
J1946+2052 & 1.495 & 0.079 & 0.06 & 0.04 & 0.002 & 1.50$\times10^{-2}$ \\
B2127+11C & 1.131 & 0.335 & 0.68 & 0.25 & 0.014 & 1.55$\times10^{-2}$ \\\hline
\multicolumn{6}{l}{NS-CO WD systems, with median WD masses above 1$\mso$ }\\
J1141-6545 & 1.213 & 0.198 & 0.17 & 0.46 & 0.026 & 1.604$\times10^{-2}$ \\
J1952+2630 & 1.133 & 0.392 & 0.00 & 3.33 & 0.235 & 2.632$\times10^{-2}$ \\\hline
\multicolumn{6}{l}{The ratio of total SFR weights of NS-WD to NS-NS: 0.23}
 \enddata
 \tablecomments{$M_{\rm comp}$ is the mass of the pulsar's companion, which is taken to be the observed median value, $P_{\rm orb}$ is the observed orbital period, $e$ is the observed eccentricity, $\tau_{\rm M}$ is the merger delay time computed by using the equation in \citet{Mandel2021RNAAS...5..223M}, $z_{\rm birth}$ is the birth redshift at a look-back time equal to the merger delay time, and SFR weight is the computed by the cosmic star formation rate density at the birth redshift by using the formula in \citet{Madau2014}.
 }
\end{deluxetable}

\section{Constant star formation\label{CSFR}}

The initial mass function $\phi_{\rm IMF}$ has the following form
\begin{equation}
    \phi_{\rm IMF} \propto M^{-\alpha} \mathrm{d}M,
\end{equation}
which gives the number fraction of stars in a certain mass interval $[M,\, M+\mathrm{d}M]$. For a constant star formation, we need to obtain the fraction of mass that is used to form stars with masses in $[M,\, M+\mathrm{d}M]$. Hence we convert the initial mass function into the mass fraction form
\begin{equation}
    \phi_{\rm IMF} \propto M^{-\alpha}\times M\, \mathrm{d}M,
\end{equation}
and the corresponding initial binary mass function is 
\begin{equation}
    \phi_{\rm IMF} \propto M^{-\alpha}(1+q)\times M\, \mathrm{d}M\,\mathrm{d}q,
\end{equation}
where $q$ is the mass ratio, and the mass of the primary star $M$ is assumed to follow the IMF.
Here we refer to the appendix in \citet{Xu2025arXiv250323876X} for detailed description.

Therefore, in the case of a constant star formation, the $f_{\rm NS-WD}$ fraction is given by 
\begin{equation}
\begin{split}
 &(f_{\rm NS-WD})_{\rm Standard} = \left(\frac{\etaWD}{\etaNS}\right)\times\left(\frac{\langle M_{\rm NS-WD} \rangle}{\langle M_{\rm NS-NS} \rangle}\right)^{-1}\times\\
 &\frac{\int_{\mnsi}^{\mbhi}\mi^{-\alpha+1}\int_{\mwdl/M_{\rm i}}^{\rm\mwdu/M_{\rm i}}(1+q_\mathrm{i})\,\dd \qi\,\dd \mi}{\int_{\mnsi}^{\mbhi}\mi^{-\alpha+1}\int_{\mwdu/M_{\rm i}}^1(1+q_\mathrm{i})\,\dd \qi\,\dd \mi}, 
 \label{eq:csfr-standard}
\end{split}
\end{equation}
where $\langle M_{\rm NS-NS} \rangle$ and $\langle M_{\rm NS-WD} \rangle$ are the average binary masses of the progenitors of NS-NS and NS-WD respectively. For a given mass range $[M_1,\,M_2]$ and mass ratio range $[q_1,\,q_2]$, the average mass $\langle M\rangle$ is evaluated by
\begin{equation}
    \langle M\rangle = \frac{\int_{M_1}^{M_2}\int_{q_1}^{q_2} M_{\rm i}^{-\alpha}\times(1+q_\mathrm{i})M_\mathrm{i}\,\dd \qi\,\dd \mi}{\int_{M_1}^{M_2}\int_{q_1}^{q_2} M_{\rm i}^{-\alpha}\,\dd \qi\,\dd \mi},
\end{equation}
where a flat mass ratio distribution is assumed, and  orbital period distribution is expected to be independent on mass ratio and initial primary mass. For our standard evolutionary path, we have 
\begin{equation}
\begin{split}
    &\langle M_{\rm NS-WD}\rangle = \\&\frac{\int_{\mnsi}^{\mbhi}\int_{\mwdl/M_{\rm i}}^{\rm\mwdu/M_{\rm i}} M_{\rm i}^{-\alpha}\times(1+q_\mathrm{i})M_\mathrm{i}\,\dd \qi\,\dd \mi}{\int_{\mnsi}^{\mbhi}\int_{\mwdl/M_{\rm i}}^{\rm\mwdu/M_{\rm i}} M_{\rm i}^{-\alpha}\,\dd \qi\,\dd \mi},
    \end{split}
\end{equation}
and 
\begin{equation}
\begin{split}
    &\langle M_{\rm NS-NS}\rangle = \\
    &\frac{\int_{\mnsi}^{\mbhi}\int_{\mwdu/M_{\rm i}}^1 M_{\rm i}^{-\alpha}\times(1+q_\mathrm{i})M_\mathrm{i}\,\dd \qi\,\dd \mi}{\int_{\mnsi}^{\mbhi}\int_{\mwdu/M_{\rm i}}^1 M_{\rm i}^{-\alpha}\,\dd \qi\,\dd \mi}
\end{split}
\end{equation}

Different from \citet{Xu2025arXiv250323876X}, in this study we only focus on the birth rate fraction of NS-WD binaries in NS-NS binaries instead of their absolute numbers. Hence, the effect of lifetime is not considered, and Eq.\,\eqref{eq:csfr-standard} can be simplified as 
\begin{equation}
\begin{split}
 &(f_{\rm NS-WD})_{\rm Standard} = \left(\frac{\etaWD}{\etaNS}\right)\times\\
 &\frac{\int_{\mnsi}^{\mbhi}\int_{\mwdl/M_{\rm i}}^{\rm\mwdu/M_{\rm i}} M_{\rm i}^{-\alpha}\,\dd \qi\,\dd \mi}{\int_{\mwdu/M_{\rm i}}^1 M_{\rm i}^{-\alpha}\,\dd \qi\,\dd \mi}, 
 \label{eq:csfr-standard-2}
\end{split}
\end{equation}
which has the same form as the star burst case adopted in the main text (Eq.\,\ref{f_NSWD}). Therefore, we conclude that our definition of birth rate ratio $f_{\rm NS-WD}$ is independent on star formation history. The main uncertainties of our estimates should lie in the details of binary evolution and the outcome merger delay times, which are captured by the $(\etaWD/\etaNS)$ factor.

\end{CJK*}
\end{document}